\documentclass[10pt,a4paper,twoside]{article}
\usepackage{epsfig}
\usepackage{baltlat6}
\usepackage{array}
\usepackage{wrapfig}
\usepackage{subfig}
\usepackage{amssymb}
\begin{document}
\ \
\vspace{0.5mm}
\setcounter{page}{1}
\vspace{8mm}

\titlehead{Baltic Astronomy, vol.\,XX, XXX--XXX, 2011}

\titleb{A COMPLEX STELLAR LINE-OF-SIGHT VELOCITY DISTRIBUTION IN THE
LENTICULAR GALAXY NGC~524}

\begin{authorl}
\authorb{I. Katkov}{1},
\authorb{I. Chilingarian}{2,1},
\authorb{O. Sil'chenko}{1},
\authorb{A. Zasov}{1} and
\authorb{V. Afanasiev}{3}
\end{authorl}

\begin{addressl}
\addressb{1}{Sternberg Astronomical Institute, Moscow State University, Universitetskii
pr. 13, Moscow, 119992 Russia;}
\addressb{2}{CDS -- Observatoire de Strasbourg, CNRS UMR 7550, Universit\'{e} de Strasbourg, 11 Rue de
l'Universit\'{e}, 67000 Strasbourg, France}
\addressb{3}{Special Astrophysical Observatory, Russian Academy of Sciences, Nizhnii Arkhyz,
Karachaevo\-Cherkesskaya Republic, 369167 Russia}
\end{addressl}

\submitb{Received: 2011 XXXXXX XX; accepted: 2011 XXXXXXX XX}

\begin{summary} 
We present the detailed study of the stellar and gaseous kinematics of the
luminous early-type galaxy NGC~524 derived from the long-slit spectroscopic
observations obtained with the Russian 6-m telescope and the IFU data from
the SAURON survey. The stellar line-of-sight velocity distribution (LOSVD)
of NGC~524 exhibits strong asymmetry. 
We performed the comprehensive analysis of the LOSVD using two complementary
approaches implemented on top of the {\sc nbursts} full spectral
fitting technique, (a) a non-parametric LOSVD recovery and (b) a parametric
recovery of two Gaussian kinematical components having different stellar
populations. We discuss the origin of the complex stellar LOSVD of NGC~524.
\end{summary}


\resthead{A complex stellar LOSVD in NGC 524}
{I.~Katkov et. al.}

\sectionb{1}{INTRODUCTION}

NGC 524 is a luminous ($M_B=-21.7$~mag) lenticular galaxy settled
in the centre of a rich group (Garcia 1993) containing a X-ray hot gas
component with a slightly lopsided distribution with respect to NGC~524
(Mulchaey et al. 2003). The galaxy demonstrates nearly circular isophotes,
$\epsilon <0.05$ (e.g. Magrelli et al. 1992), so from the photometric point
of view it looks face-on. However, kinematic measurements revealed quite
fast rotation of the galaxy (Sil'chenko 2000, Simien \& Prugniel 2000,
Emsellem et al. 2004) that is inconsistent with the photometric inclination
of $<18$~deg. Moreover, NGC~524 possesses a roundish disc of ionised gas
extending up to 4~kpc (25~arcsec) from the centre (Macchetto et al. 1996)
rotating even faster than the stars. By comparing the rotation of the
ionised gas and the stars in the centre of NGC~524, Sil'chenko (2000)
suggested that its gaseous disc was inclined with respect to the stellar
one. The recent structural analysis of NGC~524 (Sil'chenko 2009) reveals
that the bulge of this giant lenticular galaxy is rather modest and is seen
only inside $R\approx 10$~arcsec, while the general structure is dominated
by two exponential stellar discs with the scalelengths of 0.9 and 3~kpc
respectively.

%%%%%%%%%%%%%%%%%%%%%%%%%%%%%%  FIGURE1

\begin{figure}[!tH]
\vbox{
\centerline{
\psfig{figure=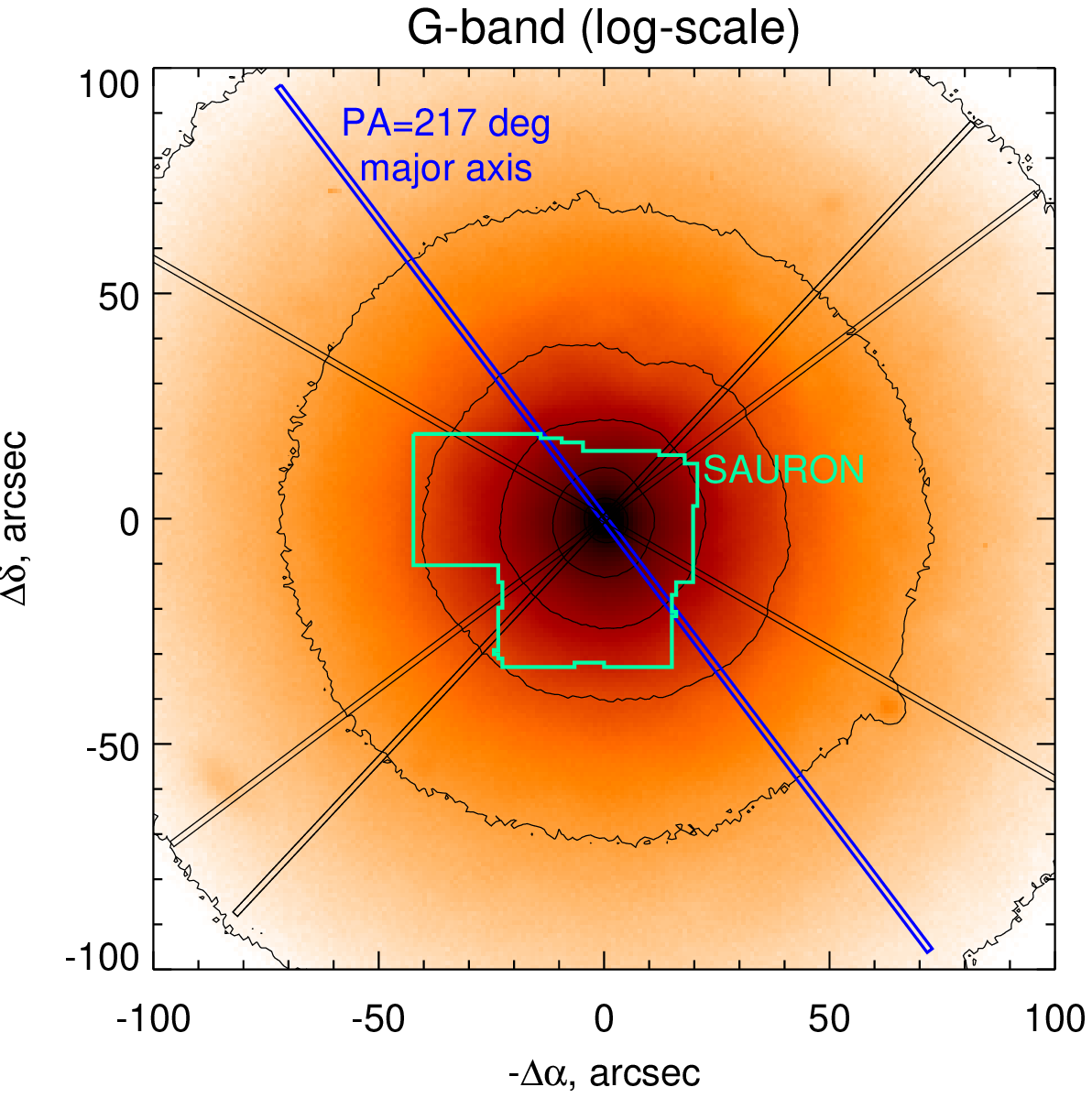,height=0.28\textheight,angle=0,clip=}
\psfig{figure=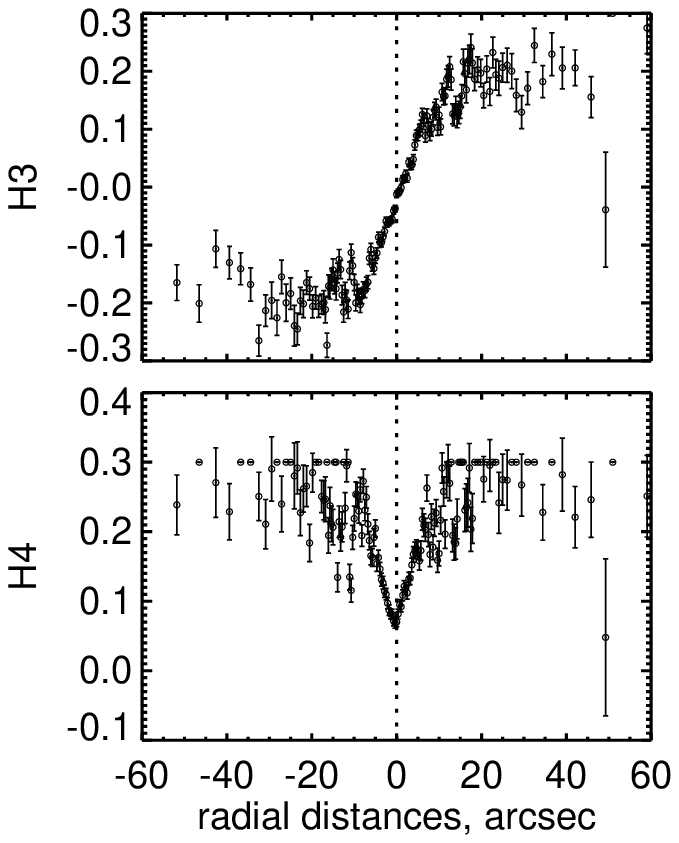,height=0.26\textheight,angle=0,clip=}
}}
\vspace{0mm}
\captionb{1}
{{\it Left:} The positions of the SCORPIO slits and the SAURON mosaic field of view are
overlapped onto the $g$-band image of NGC~524 obtained at CFHT.
{\it Right:} The Gauss-Hermite coefficients characterizing the asymmetry of the stellar LOSVD.
}
\end{figure}

%%%%%%%%%%%%%%%%%%%%%%%%%%%%%%%%%%%%%%%%%%%%%%%%%%%%%%%%%%%%%%%%%%%%%%%%%%%%%%%%%%%%%%%%%%%%%%%%%%%%%%%%%%%%%%%%%%%%%%%%%%%%%%%
\sectionb{2}{OBSERVATIONS AND DATA REDUCTION}

We used two datasets derived from long-slit and integral-field spectroscopy. 

The long-slit spectroscopic observations were obtained with the SCORPIO
universal spectrograph (Afanasiev \& Moiseev 2005) at the prime focus of the
Russian 6-m BTA telescope of the Special Astrophysical Observatory. We
observed NGC~524 in six slit positions ($P.A.=137, 217, 240, 295,
307,317$ deg, see Fig.~1 {\it left}) going through the galaxy centre. But here we present only the
long-slit spectrum obtained along the kinematical major axis
($P.A.=217$~deg). We used the ``green'' (480--550~nm) and ``red''
(610--710~nm) spectral setups covering strong stellar absorption features as
well as the emission lines of the ionised gas, $H\beta$, [O{\sc iii}]
($\lambda = 4959, 5007$~\AA), $H\alpha$, [N{\sc ii}] ($\lambda = 6548,
6583$~\AA), [S{\sc ii}] ($\lambda = 6716, 6731$~\AA) providing the spectral
resolution of 0.22~nm and 0.31~nm correspondingly.

Data reduction for the long-slit spectral data of NGC 524 was identical to
that of the lenticular galaxy NGC~7743 presented in Katkov et al. (2011a).
Briefly, the data reduction steps included: bias subtraction, flat fielding,
cosmic ray hit removal, building the wavelength solution using arc-line
spectra, constructing the spectral line spread function (LSF) variation
model using twilight spectrum and night sky spectrum subtraction taking into
account the LSF variation along and across the wavelength direction (Katkov
\& Chilingarian 2010), and adaptive binning along the slit in order to
achieve the minimal value of the signal-to-noise ratio $S/N = 20$ per
spatial bin.

We also used the data obtained with the integral-field spectrograph SAURON
(Bacon et al. 2001) at the 4.2-m William Herschel Telescope. NGC~524 was
observed in three positions of the SAURON lenslet array with $0.94$~arcsec
sampling (see Fig~1 {\it left}). The covered spectral range was 480--540~nm with
the spectral resolution 0.48~nm. For our analysis we used the science-ready
data cube kindly provided by E.~Emsellem that we binned adaptively using
Voronoi tessellation (Cappellari \& Copin 2003) to the minimal S/N ratio of
100 per bin. %The SAURON lens array positions overplotted on the galaxy image is
%shown in Fig.~1 (\emph{left}).

%%%%%%%%%%%%%%%%%%%%%%%%%%%%%%%%%%%%%%%%%%%%%%%%%%%%%%%%%%%%%%%%%%%%%%%%%%%%%%%%%%%%%%%%%%%%%%%%%%%%%%%%%%%%%%%%%%%%%%%%%%%%%%%
\sectionb{3}{DATA ANALYSIS}

\subsectionb{3.1}{SSP-EQUIVALENT PARAMETERS AND EMISSION LINE KINEMATICS}

We derived the parameters of internal kinematics and stellar populations of
NGC~524 by fitting high-resolution PEGASE.HR (Le Borgne et al. 2004) simple
stellar population (SSP) models against our spectra using the {\sc nbursts}
full spectral fitting technique (Chilingarian et al. 2007a,b). We determined
SSP-equivalent ages $T$ and metallicities $[Z/H]$ of the stellar population
as well as the stellar kinematics using the Gauss-Hermite parametrization
up-to the 4th moment, i.e. $v$, $\sigma$, $h_3$ and $h_4$ (van der Marel \&
Franx 1993). The derived parametric LOSVD exhibits a strong asymmetry
leading to the \emph{non-physical} values of $h_3$ and $h4$ (see Fig~1. {\it
right}) corresponding to significantly negative LOSVD ``wings''.

The emission line spectrum at every spatial bin was obtained by the
subtraction of the stellar contribution (i.e. the best-fitting model) from
the observed spectrum. Then we fitted it with pure Gaussians convolved with
the LSF in order to determine the kinematics of the ionised gas and emission
line ratios.

%%%%%%%%%%%%%%%%%%%%%%%%%%%%%%  FIGURE2

\begin{figure}[!tH]
\vbox{
\centerline{
\psfig{figure=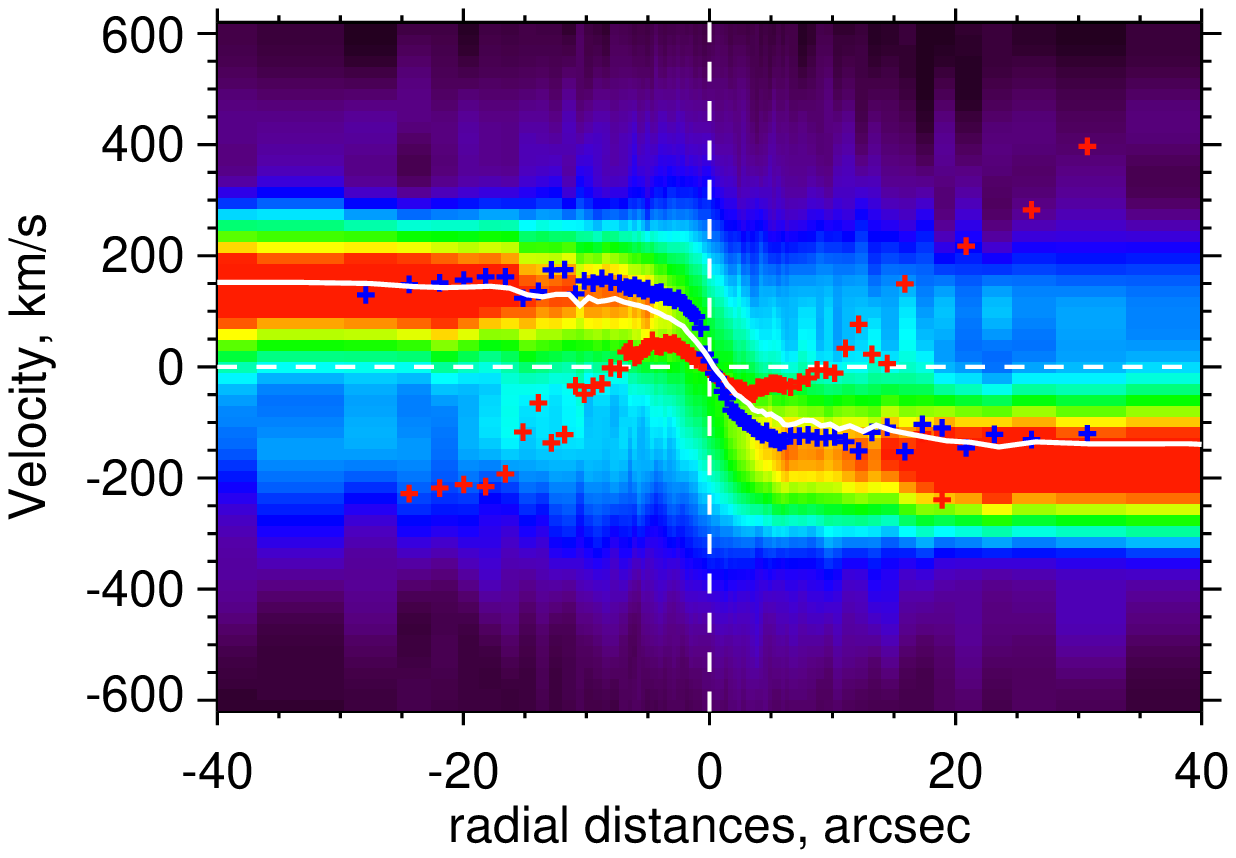,height=0.23\textheight,angle=0,clip=}
\psfig{figure=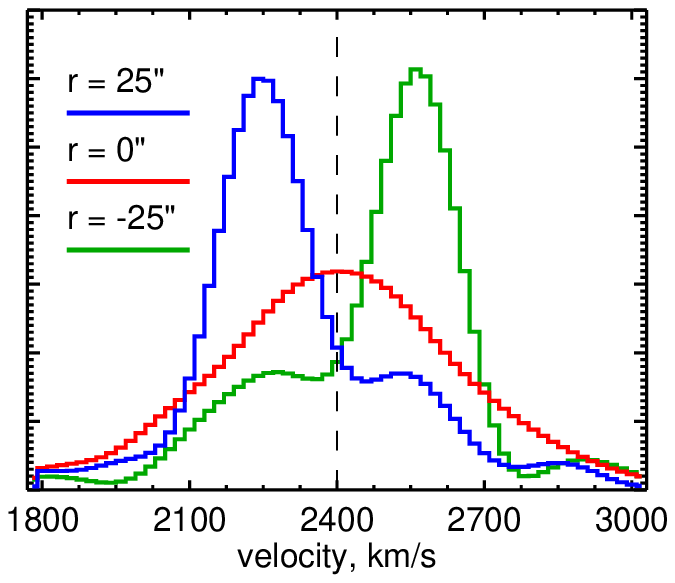,height=0.23\textheight,angle=0,clip=}
}}
\vspace{1mm}
\captionb{2} 
{{\it Left:} The position-velocity diagram. White line presents a radial velocity profile obtained using one-component model. Red and blue crosses - ''bulge'' and ''disc'' components in the two-component model. {\it Right:} The LOSVD at radii at -25,0,25 arcsec.}
\end{figure}

\subsectionb{3.2}{NON-PARAMETRIC LOSVD}

We propose a non-parametric recovery technique based on the full spectral
fitting requiring no \emph{a priory} LOSVD knowledge. The logarithmically
rebinned model spectrum, $\mathcal{F}(\lambda)$ is the convolution of the assumed
normalized LOSVD, $\mathcal{L}(v)$, with the rest-frame SSP model, $\mathcal{F}_r(w)$:
 $$     \mathcal{F}(w) = \int_{u_{min}}^{u_{max}} \mathcal{F}_r(w-u)\mathcal{L}(u)
du,$$
where $w=\ln(\lambda)$, $u=\ln(1+v/c)$. We used the output SSP model of
the {\sc nbursts} fitting as a template spectrum $\mathcal{F}_r$. This equation
 can be considered as a linear inverse problem whose
solution is very sensitive to the noise in the data. Hence, we chose to
regularize the problem be requiring the LOSVD to be smooth. In order to do so,
we use the cubic penalization
$\mathcal{P}(\mathcal{L})=\mathcal{L}^T\cdot\mathcal{D}^T\cdot\mathcal{D}\cdot\mathcal{L}$,
where $\mathcal{D}$ - is the third-order difference operator. The function
to be minimized is given by $\chi^2 + \lambda \mathcal{P}(\mathcal{L})$. For
discussion on the choice of $\lambda$ see Press et al. (2007).

Using this technique, we confirmed a strong asymmetry of the NGC~524 LOSVD.
Fig.~2 displays the result of the LOSVD recovery for the long-slit data
along the kinematical major axis as a position-velocity diagram ({\it left}).
The LOSVD profiles at radii $-25$, $0$ and $25$~arcsec are shown to the right.

%%%%%%%%%%%%%%%%%%%%%%%%%%%%%%  FIGURE3
\begin{figure}
  \centerline{
      \psfig{figure=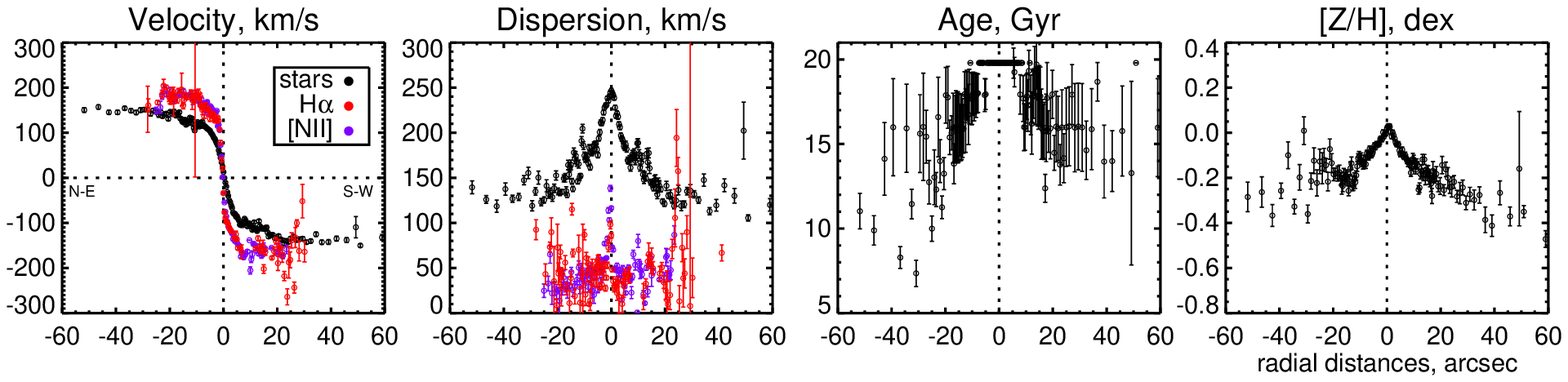,width=0.95\textwidth,angle=0,clip=}}
\centerline{
      \psfig{figure=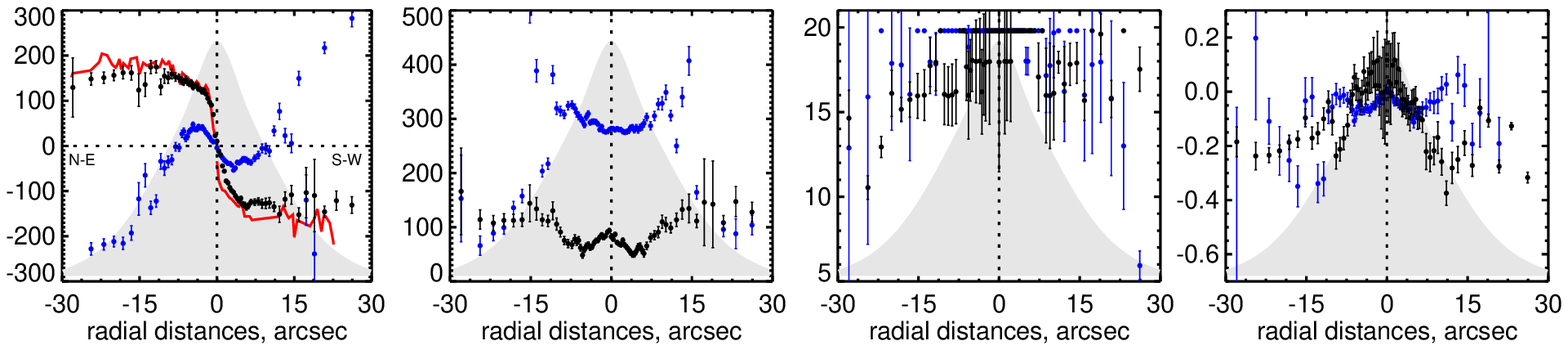,width=0.95\textwidth,angle=0,clip=}}
  \captionb{3}{SCORPIO long-slit spectroscopic data. First row -- one-component spectra
fitting with S/N ratio equal to 20 in the green spectral domain. Second row 
-- two component spectra fitting in green domain ($S/N=40$). Black points correspond to ''disc'' component, blue points to ''bulge'' one, red line shows $H\alpha$ kinematics.} 
\end{figure}

\subsectionb{3.3}{TWO-COMPONENT PARAMETRIC LOSVD RECOVERY}

Another approach we use is a full spectral fitting using 
a two-component model where different stellar population components  
have two different pure Gaussian LOSVDs. An optimal template
is represented by the linear combination of two SSPs each convolved with its
own LOSVD, hence the $\chi^2$ value is computed as follows:
$$
 \chi^2 = \sum_{N_\lambda}\frac{ ( F_i - P_p\cdot \sum_{j=1}^{j=2}k_j \cdot
S(T_j,Z_j) \otimes \mathcal{L}(v_j,\sigma_j) )^2}{\delta F_i^2},
$$
where $\mathcal{L}(v,\sigma)$ - pure Gaussian LOSVD; $F_i$ and $\delta F_i$
are observed flux and its uncertainty; $S(T_j,Z_j)$ is the flux from the
$j$-th synthetic spectrum of SSP with given age $T_j$ and metallicity $Z_j$;
$P_p$ is multiplicative Legendre polynomials of order $p$ for correcting the
continuum which determined at each step of minimization loop by solving the
linear least-square problem; $k_j$ is the $j$-th component weight (normally
found by the linear minimization). The important point in this study is that
we fixed the relative SSP contributions $k_j$ to the values derived from the
light profile decomposition.

%%%%%%%%%%%%%%%%%%%%%%%%%%%%%%%%%%%%%%%%%%%%%%%%%%%%%%%%%%%%%%%%%%%%%%%%%%%%%%%%%%%%%%%%%%%%%%%%%%%%%%%%%%%%%%%%%%%%%%%%%%%%%%%

\sectionb{4}{RESULTS AND DISCUSSION}

\subsectionb{4.1}{KINEMATICS}

At all radii the LOSVD of NGC~524 clearly demonstrates the presence of at
least two components clearly visible in Figs.~1--2 confirmed independently
by the non-parametric and parametric LOSVD reconstruction on SAURON and
SCORPIO datasets. The inner region of the galaxy that corresponds to the
small exponential pseudobulge identified by Sil'chenko (2009) is very hot in
a dynamical sense with the velocity dispersion exceeding 300~km~s$^{-1}$.

This component disappears at about $R \approx 20$~arcsec, where we see a
sharp drop in the velocity dispersion and radial velocities increasing
outwards suggesting the counter-rotation with respect to the main disc (see
blue profiles in the bottom panels of Fig.~3). At large radii
($R>20$~arcsec), the LOSVD recovered non-parametrically also becomes
strongly bi-modal with the secondary component clearly visible in Fig.~1.

%%%%%%%%%%%%%%%%%%%%%%%%%%%%%%  FIGURE4

\begin{figure}[!tH]
\vbox{
\centerline{
\psfig{figure=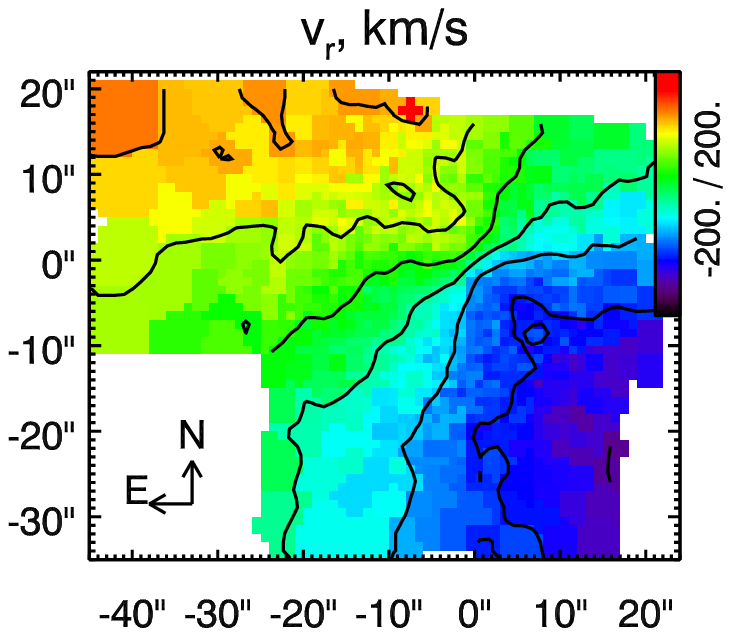,width=0.25\textwidth,angle=0,clip=}
\psfig{figure=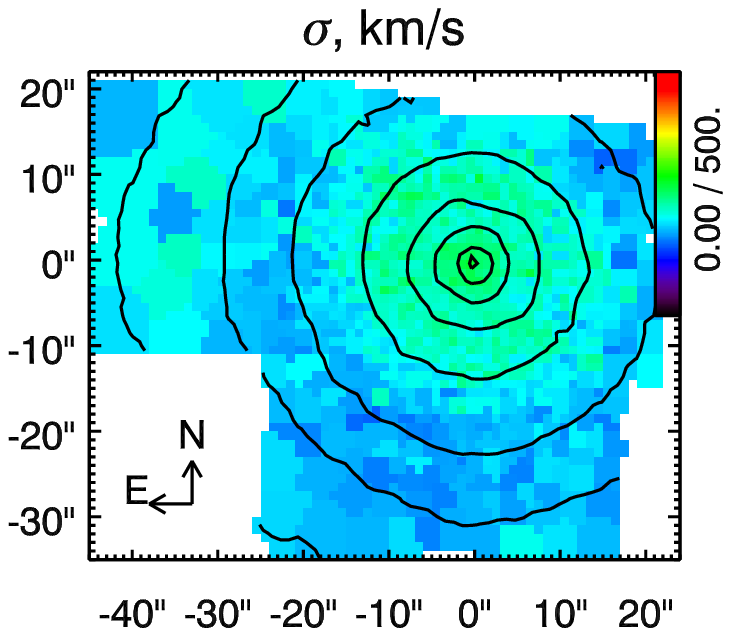,width=0.25\textwidth,angle=0,clip=}
\psfig{figure=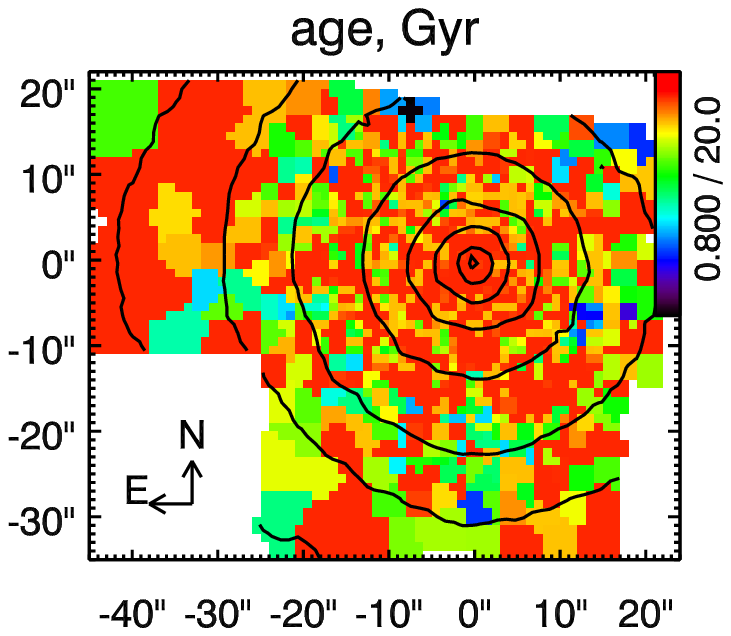,width=0.25\textwidth,angle=0,clip=}
\psfig{figure=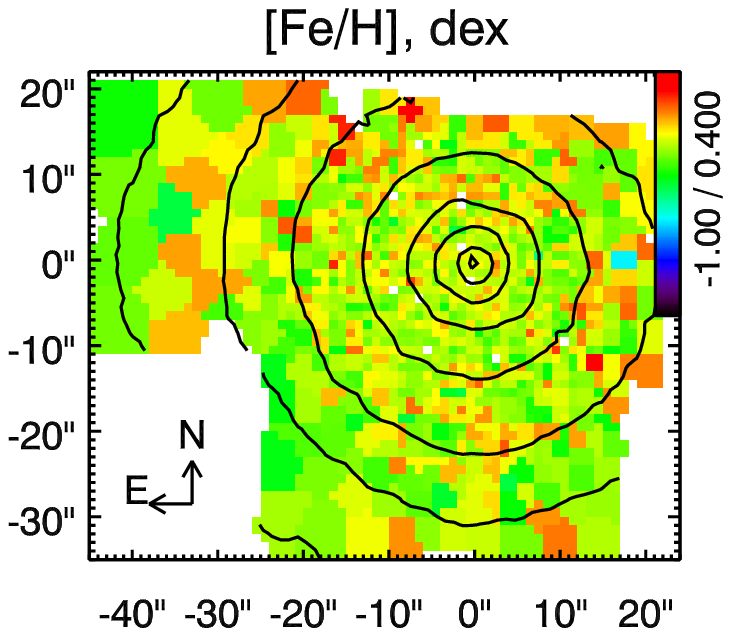,width=0.25\textwidth,angle=0,clip=}
}
\centerline{
\psfig{figure=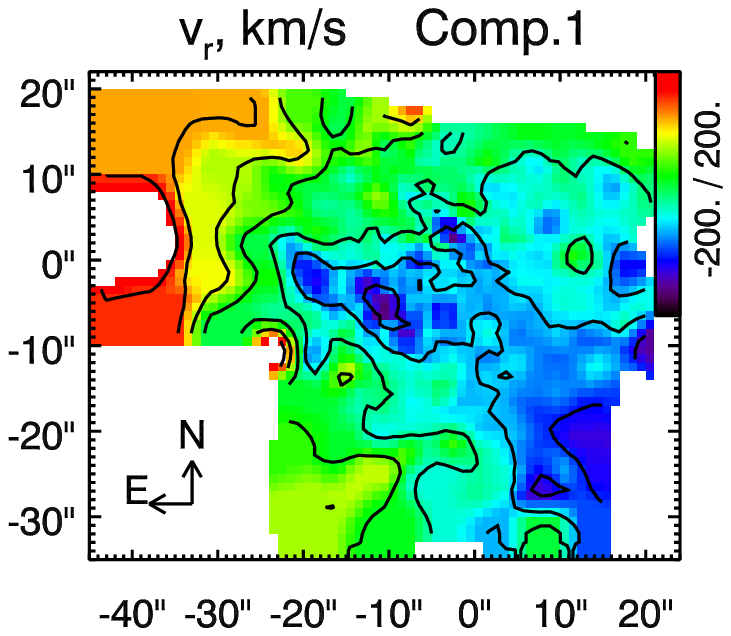,width=0.25\textwidth,angle=0,clip=}
\psfig{figure=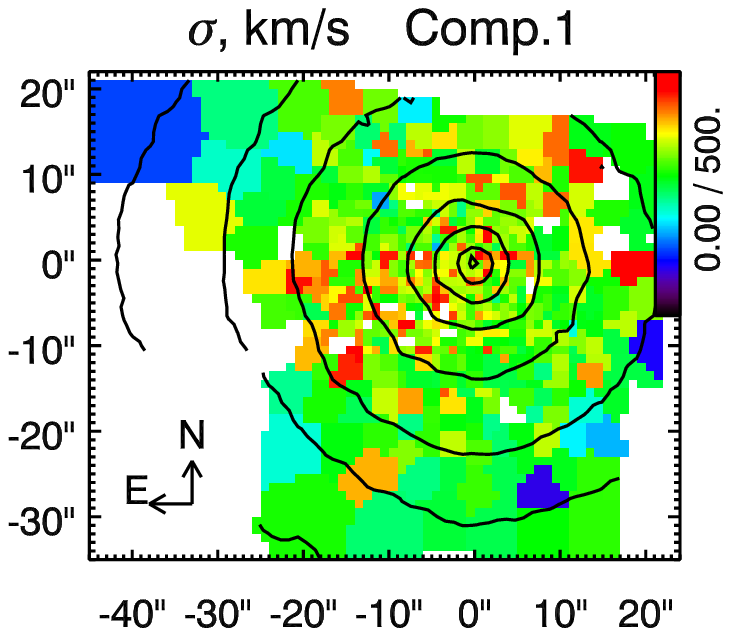,width=0.25\textwidth,angle=0,clip=}
\psfig{figure=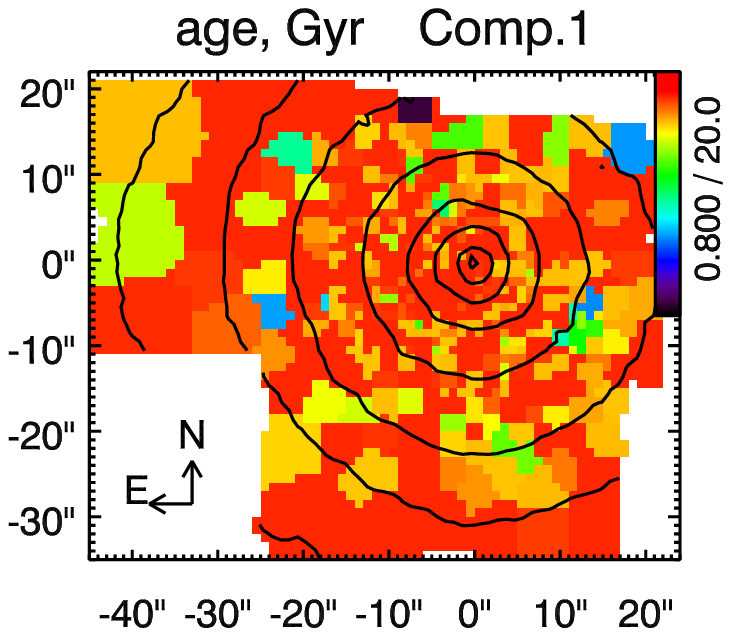,width=0.25\textwidth,angle=0,clip=}
\psfig{figure=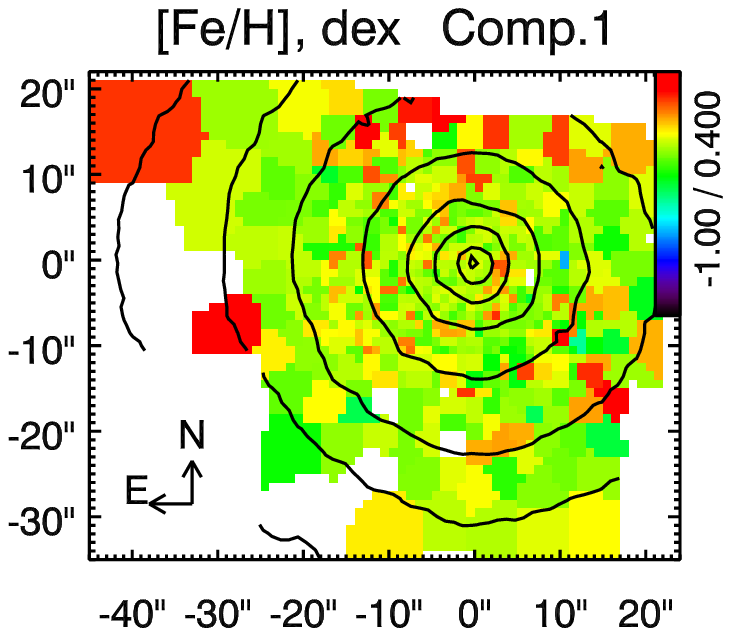,width=0.25\textwidth,angle=0,clip=}
}
\centerline{
\psfig{figure=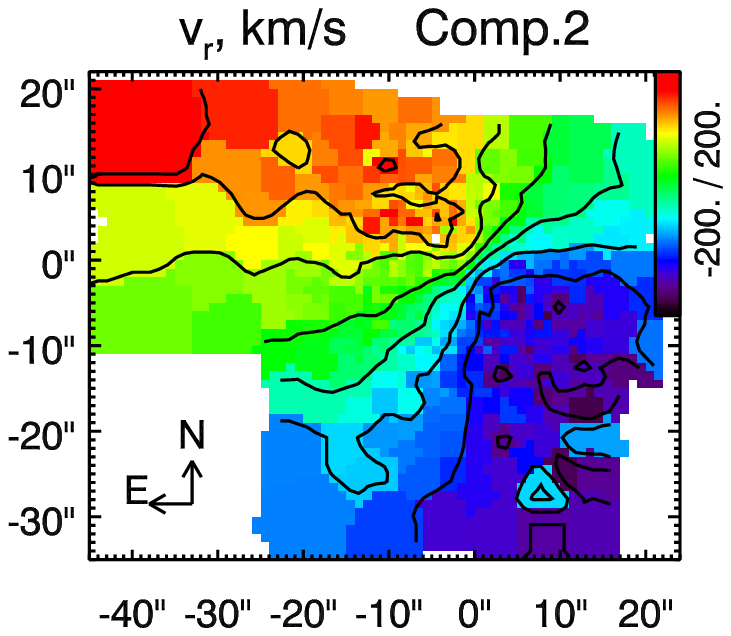,width=0.25\textwidth,angle=0,clip=}
\psfig{figure=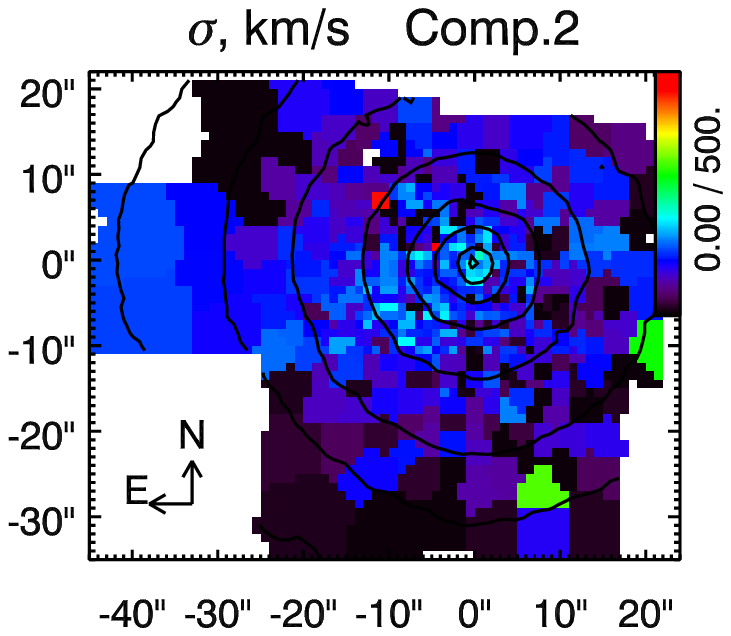,width=0.25\textwidth,angle=0,clip=}
\psfig{figure=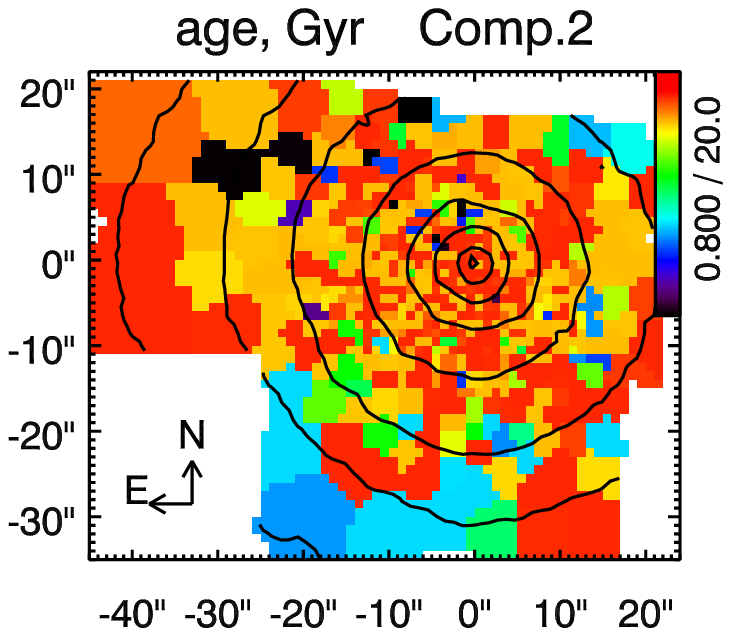,width=0.25\textwidth,angle=0,clip=}
\psfig{figure=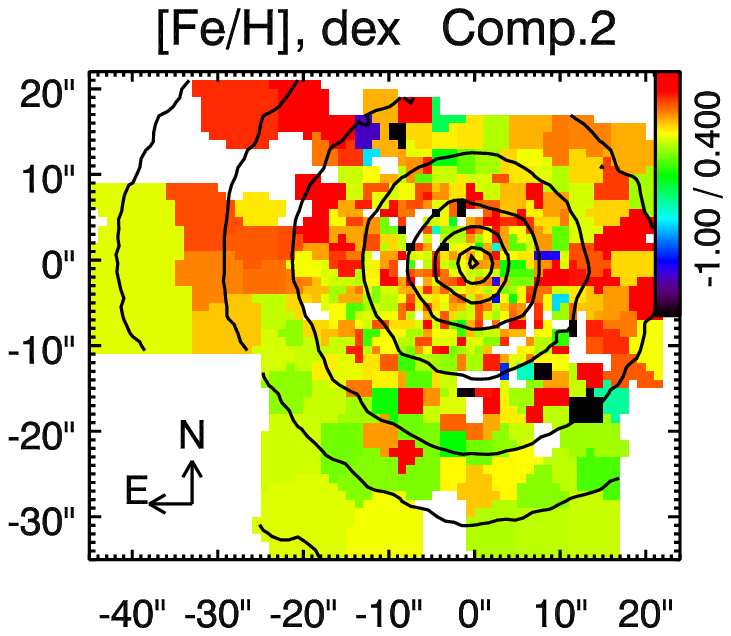,width=0.25\textwidth,angle=0,clip=}
}}
\vspace{1mm}
\captionb{4}
{SAURON integral-field spectroscopic data. From left to right: velocity, velocity dispersion, SSP-equvalent ages and metallicities. First row correspond to one component fitting spectra. Second and third rows present a ''bulge'' and ''disc'' parameter maps for two component spectra decomposition.}
\end{figure}

\subsectionb{4.2}{STELLAR POPULATIONS AND EMISSION LINE RATIOS}

The radial variations of the stellar population parameters derived from
the SSP fitting of the long-slit data are shown in Fig.~3
(right-top). The mean stellar age is very old ($T \gtrsim 15$~Gyr) being at the limit of
the SSP model grid in the central 20~arcsec. The metallicity in the centre
is close to the Solar value decreasing down to $[Z/H]\approx-0.4$~dex in
the outer disc. The results of spectrum fitting using a two-component
model are presented in Fig.~3 (right-bottom). ``Bulge'' (blue points) and ``disc''
(black points) components do not have significant differences of the age
radial profile being both very old with the ``bulge'' reaching the limiting
age in the models. The ``disc'' metallicity decreases from
$[Z/H]\approx0.1$~dex in the centre to -0.2~dex at large radii. 
Our analysis of the virtually noiseless ($S/N > 100$) SAURON
dataset (see Fig.~4, right columns) agree very well with the long-slit
data. The mean stellar parameters of the ``bulge'' component are $T>15$ Gyr
and $[Z/H]\approx-0.1$~dex, while for the ``disc'' component they
are $T\approx 14$~Gyr and $[Z/H]\approx 0.1$~dex.

The analysis of emission line ratios using the classical diagnostics
[O{\sc iii}]/H$\beta$ vs [N{\sc ii}]/H$\alpha$ (Baldwin et al. 1981) rules
out the ongoing star formation in the disc of NGC~524 leaving the space for
the shockwave excitation or the nuclear activity. However, the latter
mechanism cannot explain the large spatial extent of the emission line
region.

\subsectionb{4.3}{DISCUSSION}

A very good agreement between the dynamically cold ``disc'' component of the
stellar LOSVD and the ionised gas kinematics suggest that the gas is
rotating in the plane of the main stellar disc. A small discrepancy of the
rotation (20~km~s$^{-1}$) can be explained by the asymmetric drift and well
corresponds to the observed stellar velocity dispersion of the disc
component of $\sim100$~km~s$^{-1}$. Old stellar population and
the emission-line diagnostics rule out both, recent and ongoing star
formation in the gaseous disc. The second kinematical component 
(blue profiles in Fig.~3) probably corresponds to two different structures
at different radii. The dynamically hot inner part without much rotation is
a manifestation of the compact central pseudo-bulge, while at $R>20$~arcsec
we see the presence of a counter-rotating disc component that is
supported by the drops in the velocity dispersion (300 to
$<$100~km~s$^{-1}$) and metallicity (0.0 to $-0.3$~dex) profiles.

The origin of NGC~524 has to be investigated in detail using
state-of-art numerical simulations. Right now we can speculate about its
evolution based on the observational results we have. NGC~524 might have
originated from the face-on collision of two initially counter-rotating
co-planar giant disc galaxies. The metallicity difference between the two
components suggests the merger mass ratio of about 1:4. The gas in the main
stellar disc of NGC~524 might have survived from the original galaxy or
collected later from mergers with low-mass satellites or from the accretion
from the cosmic filaments. However, its surface density is still below the 
threshold and therefore it prevents the start of the star formation.

\References

\refb Afanasiev, V. L., Moiseev, A. V. 2005, Astronomy Letters, 31, 193
\refb Bacon, R., Copin, Y., Monnet, G. et al. 2001, MNRAS, 326, 23
\refb Baldwin, J.~A., Phillips M.~M., Terlevich R., 1981, PASP, 93, 5
\refb Cappellari, M., Copin, Y. 2003, MNRAS, 342, 345
\refb Chilingarian, I.~V., Prugniel, P., Sil'chenko, O.~K., Afanasiev, V.~L. 2007a, MNRAS, 376, 1033
\refb Chilingarian, I.~V., Prugniel, P., Sil'chenko, O.~K., Koleva, M.
2007b, In: Stellar Populations as Building Blocks of Galaxies, Proc. of the IAU
Symp. no 241, eds. A. Vazdekis and R. F. Peletier, p. 175, arXiv:0709.3047
\refb Emsellem E. et al. 2004, MNRAS 352, 721
\refb Garcia A.M. 1993, A\& A Suppl. Ser. 100, 47
\refb Katkov, I.Yu., Moiseev, A.V., Sil'chenko, O.K. 2011a ApJ, submitted
\refb Katkov, I.Yu., Chilingarian I.V. 2011,  arXiv:1012.4125
\refb Le Borgne, D., Rocca-Volmerange, B., Prugniel, et al. 2004, A\&A, 425, 881
\refb Macchetto F., Pastoriza M., Caon N., Sparks W. B., Giavalisco M., Bender R., 
         Capaccioli M. 1996,  A\& A Suppl. Ser. 120, 463
\refb van der Marel, R.P. \& Franx, M. 1993, ApJ, 407, 525
\refb Magrelli G., Bettoni D., Galletta G. 1992, MNRAS 256, 500
\refb Mulchaey J., Davis D., Mushotzky R., Burstein D. 2003, ApJ Suppl 145, 39
\refb Press, W.H., Teukolsky S.A., Vetterling, W.T., Plannery, B.P. 2007, Numerical Recipes: The Art of Scientific Computing, 3d edn, Cambridge Univ. Press %, Cambridge
\refb Sil'chenko O.K. 2000, AJ 120, 741
\refb Sil'chenko O.K. 2009, In: The Galaxy Disk in Cosmological Context, Proc. of the IAU 
       Symp. no 254, eds. J. Andersen, J. Bland-Hawthorn, and B. Nordstrom, p.~173
\refb Simien F., Prugniel Ph. 2000,  A\& A Suppl. Ser. 145, 263

\end{document}